\documentclass[10pt]{article}

\usepackage{amsfonts,amsmath,amssymb}
\usepackage{mathtools}
\usepackage{stackrel}
\usepackage{leftidx}
\usepackage[amsmath,amsthm,thmmarks]{ntheorem}

\usepackage{authblk}

\usepackage{euscript}
\usepackage{pxfonts}
\usepackage{dsfont}
\usepackage{bbold}

\usepackage[shortlabels]{enumitem}

\usepackage{hyperref}
\usepackage{cleveref}

\usepackage{xcolor}
\usepackage{tcolorbox}
\usepackage{graphicx}
\usepackage{subfigure}

\usepackage[all,2cell,arrow,matrix]{xy} \UseAllTwocells \SilentMatrices
\usepackage{tikz-cd}
\usepackage{tikz}
\usetikzlibrary{arrows,arrows.meta,decorations.markings,decorations.pathreplacing,calc}

\usepackage{verbatim}
\usepackage{comment}

\usepackage{epstopdf} 

\usepackage[nottoc]{tocbibind} 
\usepackage{appendix}

\newcommand\void[1]       {}

\tikzset{->-/.style={decoration={markings,mark=at position #1 with {\arrow{stealth}}},postaction={decorate}},->-/.default=0.55}
\colorlet{e_ext}{red}
\colorlet{m_ext}{blue!30}
\tikzset{e_str/.style={very thick,red!80}}
\tikzset{m_str/.style={very thick,blue!80}}
\tikzset{m_dual_str/.style={thick,dashed,blue}}
\tikzset{link_label/.style={scale=0.8,black}}

\theoremstyle{definition}

\newtheorem{thm}{Theorem}[subsection]

{
\theoremsymbol{\mbox{${\blacksquare}$}}
\newtheorem{defn}[thm]{Definition}
}
{
\theoremsymbol{\mbox{$\vardiamondsuit$}}
\newtheorem{expl}[thm]{Example}
}
{
\theoremsymbol{\mbox{$\diamondsuit$}}
\newtheorem{rem}[thm]{Remark}
}
\qedsymbol{\mbox{$\square$}}

\numberwithin{equation}{subsection}

\newcommand\be            {\begin{equation}}
\newcommand\ee            {\end{equation}}
\newcommand\bea           {\begin{eqnarray}}
\newcommand\eea         {\end{eqnarray}}
\newcommand\bnu          {\begin{enumerate}}
\newcommand\enu          {\end{enumerate}}
\newcommand\bit          {\begin{itemize}}
\newcommand\eit          {\end{itemize}}

\newcommand{\pf}{\begin{proof}}
\newcommand{\epf}{\qed\end{proof}}

\makeatletter
\providecommand{\leftsquigarrow}{%
  \mathrel{\mathpalette\reflect@squig\relax}%
}
\newcommand{\reflect@squig}[2]{%
  \reflectbox{$\m@th#1\rightsquigarrow$}%
}
\makeatother

\DeclareMathAlphabet{\mathcal}{OMS}{cmsy}{m}{n}	
\DeclareMathAlphabet{\mathsf}{OT1}{cmss}{m}{n}	

\newcommand\Cb			{\mathbb{C}}

\newcommand\Zb			{\mathbb{Z}}

\newcommand\CA			{\EuScript{A}}
\newcommand\CB			{\EuScript{B}}
\newcommand\CC			{\EuScript{C}}
\newcommand\CD			{\EuScript{D}}

\newcommand\CF			{\EuScript{F}}
\newcommand\CG			{\EuScript{G}}

\newcommand\CM			{\EuScript{M}}
\newcommand\CN			{\EuScript{N}}

\newcommand\CP			{\EuScript{P}}
\newcommand\CQ			{\EuScript{Q}}
\newcommand\CR			{\EuScript{R}}

\newcommand\CX			{\EuScript{X}}

\newcommand{\FZ}			{\text{\usefont{U}{euf}{m}{n}Z}}

\DeclareMathOperator{\id}{id}

\DeclareMathOperator{\Fun}{\EuScript{F}\mathrm{un}}

\newcommand{\one}			{\mathbb{1}}
\newcommand{\Ob}          {\mathrm{Ob}}

\newcommand\vect			{\mathrm{Vec}}

\newcommand\arXiv[1]{\href{http://arxiv.org/abs/#1}{arXiv:#1}}

\begin{document}

\begin{center} \LARGE
Categorical computation
\end{center}

\vskip 1em
\begin{center}
{\large
Liang Kong$^{a,b,c}$,\,
Hao Zheng$^{a,c,d,e,f}$\,
~\footnote{Emails:
{\tt  kongl@sustc.edu.cn, haozheng@mail.tsinghua.edu.cn}}}
\\[1em]
$^a$ Shenzhen Institute for Quantum Science and Engineering,\\
Southern University of Science and Technology, Shenzhen 518055, China 
\\[0.7em]
$^b$ International Quantum Academy, Shenzhen 518048, China \\[0.7em]
$^c$ Guangdong Provincial Key Laboratory of Quantum Science and Engineering, \\
Southern University of Science and Technology, Shenzhen 518055, China 
\\[0.7em]
$^d$ Institute for Applied Mathematics, 
Tsinghua University, Beijing 100084, China 
\\[0.7em]
$^e$ Beijing Institute of Mathematical Sciences and Applications, Beijing 101408, China
\\[0.7em]
$^f$ Department of Mathematics, Peking University, Beijing 100871, China
\end{center}

\vskip 2em

\begin{abstract}
In quantum computing, the computation is achieved by linear operators in or between Hilbert spaces. In this work, we explore a new computation scheme, in which the linear operators in quantum computing are replaced by (higher) functors between two (higher) categories. If from Turing computing to quantum computing is the first quantization of computation, then this new scheme can be viewed as the second quantization of computation. The fundamental problem in realizing this idea is how to realize a (higher) functor physically. We provide a theoretical idea of realizing (higher) functors physically based on the physics of topological orders.
\end{abstract}

\section{Classical and quantum computation} \label{sec:introduction}

Classical computation or Turing computation can be summarized mathematically as the computation of functions between sets. More precisely, for any positive integers $n$, and  
\[ \label{eq:function}
\mbox{a function}\,\,\,\,\, f: \{0,1\}^n \to \{0,1\}^n,
\] 
we design a finite set of gates $S$, each of which is also a function between two sets, such that $f$ can be realized by a circuit constructed from the gates in $S$. The classical computer is physically possible because we can realize or simulate a function between two sets physically through a circuit consisting of a family of more elementary physically realizable functions (i.e. gates). 

\medskip
The quantum computation can be summarized mathematically as the computation of linear maps (or linear operators). More precisely, for any positive integers $n$, and  
\[ \label{eq:linear-map}
\mbox{a linear map}\,\,\,\,\,F: (\Cb^2)^{\otimes n} \to (\Cb^2)^{\otimes n}, 
\]
we design a finite set of gates, each of which is also a linear map between two finite dimensional Hilbert spaces, such that $F$ can be physically realized by a quantum circuit of gates. A quantum computer is physically possible because we can physically realize linear operators in Hilbert spaces. A quantum computation is a physical way of realizing parallel computation via quantum mechanics. The idea is to replace the bit $\{ 0,1\}$ by a qubit $\Cb^2$, which can be viewed as a linear span of two states $|0\rangle$ and $|1\rangle$. Different from $\{ 0,1\}$, the states in a qubit are much more than $|0\rangle$ and $|1\rangle$. A generic state $|a\rangle$ is a superposition of them, i.e. 
\[
|a\rangle = a_0|0\rangle + a_1|1\rangle = \left( \begin{array}{c} a_0 \\ a_1\end{array} \right), \quad \forall a_1,a_2\in\Cb.
\] 
It means that $|a\rangle$ can be in either the state $|0\rangle$ or the state $|1\rangle$ with potentially non-trivial probabilities. 
Using $|0\rangle$ and $|1\rangle$, a linear map $F: \Cb^2 \to \Cb^2$ can be expressed as a matrix.  
\[
F: \,\,\, \left( \begin{array}{c} a_0 \\ a_1\end{array} \right) \,\, \mapsto \,\,  
\left( \begin{array}{cc} F_{00} & F_{01} \\ F_{10} & F_{11} \end{array} \right)  \left( \begin{array}{c} a_0 \\ a_1\end{array} \right)
= \left( \begin{array}{c} F_{00}a_0+F_{01}a_1 \\ F_{10}a_0 + F_{11}a_1\end{array} \right). 
\]
This linear operator $F$ that realizes the parallel computation of both $F(|0\rangle)$ and $F(|1\rangle)$ at the same time can be viewed as a probabilistic circuit (or quantum circuit), which can be simulated by a family of more elementary linear operators called quantum gates.

\section{The idea of categorification}

The idea of categorification was originated from the second quantization in physics. Second quantization has a long history. It is a way to obtain higher dimensional quantum field theories (QFT) from quantum mechanics, which can also be viewed as a 0+1D (spacetime dimension) QFT. 
In the early 90's of the 20th century, Luise Crain and Igor Frenkel introduced the idea of categorification aiming at constructing higher dimensional QFTs \cite{CF94}. Roughly speaking, the categorification can be viewed as a mathematical formulation of the second quantization. The process of categorification lifts lower categorical structures to higher categorical structures. 
The idea of categorification that we need in this work is rather simple and can be summarized by the following diagram: 
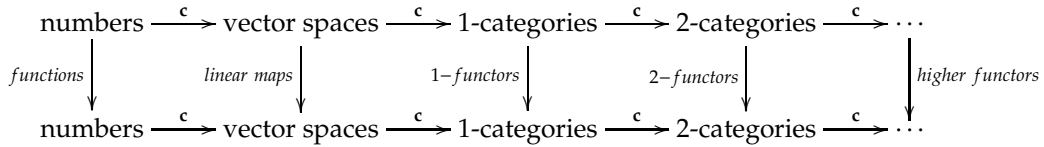
\[
\xymatrix{
\mbox{numbers} \ar[r]^-{\mathbf{c}} \ar[d]_{functions} & \mbox{vector spaces} \ar[r]^-{\mathbf{c}} \ar[d]_{linear\,\, maps} & \mbox{$1$-categories} \ar[r]^-{\mathbf{c}} \ar[d]_{1-functors} & \mbox{$2$-categories} \ar[r]^-{\mathbf{c}} \ar[d]_-{2-functors} & \cdots  \ar[d]^{higher \,\, functors} \\
\mbox{numbers} \ar[r]^-{\mathbf{c}} & \mbox{vector spaces} \ar[r]^-{\mathbf{c}} & \mbox{$1$-categories} \ar[r]^-{\mathbf{c}} & \mbox{$2$-categories} \ar[r]^-{\mathbf{c}} & \cdots
}
\]
where we use ``$\mathbf{c}$'' to represent the process of categorification. 

\begin{rem}
The reversed process is called de-categorification and can be realized by computing a generalized notion of `dimension' also called $K_0$-group in mathematics (see \cite{Wei13} for a recent review). For example, the $K_0$-group of a 2-category is a 1-category, that of a 1-category is a vector space, and that of a vector space is its dimension (i.e. a number). 
\end{rem}

Therefore, we see that if we categorify the usual quantum computation, we should obtain the computation of $n$-functors between $n$-categories. This new computation can be called a second quantized computation or a categorified quantum computation, or perhaps, even better an {\it $n$-categorical computation}. 

\medskip
Now we explain why a 1-categorical computation can be viewed as a parallel quantum computation. We recall the definition of a $1$-category. For the convenience of physical applications, 1-categories and 1-functors are all assumed to be $\Cb$-linear. 
\begin{defn}
A $1$-category $\CC$ consists of 
\bnu
\item a set of objects $\Ob(\CC)=\{ a, b, c, \cdots \}$; (We often use $a,b,c\in \CC$ for simplicity.)
\item a vector space $\hom_\CC(a,b)$ over $\Cb$ for each orders pair $(a,b)$ and $a,b\in \CC$; (a vector $f\in \hom_\CC(a,b)$ is called a morphism from $a$ to $b$ and is often denoted by $a\xrightarrow{f} b$ or $f: a\to b$).  
\item a distinguished vector $1_a \in \hom_\CC(a,a)$; 
\item a $\Cb$-linear composition map $\circ$ for $a,b,c\in \CC$: 
\begin{align*}
\hom_\CC(b,c) \otimes_\Cb \hom_\CC(a,b) &\xrightarrow{\circ} \hom_\CC(a,c) \\
g \otimes_\Cb f &\mapsto g\circ f; 
\end{align*}
\enu
satisfying the following conditions: 
\bnu
\item $(h\circ g) \circ f = h\circ (g\circ f)$ for all $a\xrightarrow{f} b \xrightarrow{g} c \xrightarrow{h} d$; 
\item $f\circ 1_a = f$ and $1_b \circ f = f$ for all $f:a\to b$. 
\enu
\end{defn}

If readers encounter the notion of a category for the first time, it is useful to keep in mind the following picture of the basic structures in a category. 
\be \label{diag:category-abc}
\xymatrix{
a \ar@(ul,ur)[]^{\hom_\CC(a,a)}  \ar@/^/[rrr]^{\hom_\CC(a,b)} & & & b \ar@(ul,ur)[]^{\hom_\CC(b,b)} \ar@/^/[lll]^{\hom_\CC(b,a)} \ar@/^/[rrr]^{\hom_\CC(b,c)} 
& & & c \ar@(ul,ur)[]^{\hom_\CC(c,c)}  \ar@/^/[lll]^{\hom_\CC(c,b)}
} 
\ee

\begin{expl}
We provide three most useful examples of 1-categories. 
\begin{enumerate}
\item $\vect$ is the 1-category of finite dimensional vector spaces over $\Cb$. Its objects $a,b,c, \cdots$ are finite dimensional vector spaces, and $\hom_\vect(a,b)$ are precisely the space of all the linear maps from $a$ to $b$. The composition maps are the usual composition of linear maps, and the identity morphism $1_a: a\to a$ is the identity linear map. 

\item $\mathrm{Rep}(G)$ is the 1-category of finite dimensional representations of a finite group $G$. More precisely, its objects are finite dimensional representations of $G$ and 1-morphisms are linear maps that intertwine the $G$-actions. The compositions and the identity morphisms are the same as those in $\vect$. 

\item $\vect_G$ is the 1-category of finite dimensional $G$-graded vector spaces for a finite group $G$. As a $\Cb$-linear 1-category, it is a direct sum of $|G|$-copies of $\vect$, i.e. $\vect_G=\oplus_{g\in G} \vect_g$. 

\end{enumerate}

\end{expl}

\begin{rem} \label{rem:n-cat}
An $n$-category $\CC$ can be described by the same diagram in (\ref{diag:category-abc}) except that the hom space $\hom_\CC(a,b)$ now becomes an $(n-1)$-category and the composition map $\circ$ now becomes an $(n-1)$-functor (see Remark\,\ref{rem:n-functor}). 
\end{rem}

\begin{defn}
A 1-functor $\CF: \CC \to \CD$ between two 1-categories consists of 
\bnu
\item a map $\CF: \Ob(\CC) \to \Ob(\CD)$ (we slightly abuse the notation here); 
\item a linear map $\CF_{a,b}: \hom_\CC(a,b) \to \hom_\CD(\CF(a), \CF(b))$ for all $a,b\in\CC$;
\enu
satisfying the following conditions: 
\bnu
\item $\CF_{a,a}(1_a)=1_{\CF(a)}$, for all $a\in\CC$; 
\item $\CF_{b,c}(g) \circ \CF_{a,b}(f) = \CF_{a,c}(g\circ f)$ for all $f\in\hom_\CC(a,b), g\in\hom_\CC(b,c)$. 
\enu
\end{defn}

We provide an intuitive picture of such a 1-functor $\CF$ by the following diagram: $\forall a,b,c\in\CC$, 
\be \label{diag:1-functor}
\raisebox{2.5em}{\xymatrix@R=1em{
a \ar[rr]_{\hom_\CC(a,b)} \ar@{|->}[dd] &  & b \ar@{|->}[dd] \ar[rr]_{\hom_\CC(b,c)} &  & c \ar@{|->}[dd] \\
& \downarrow \CF_{a,b} & &  \downarrow \CF_{b,c} &  \\
\CF(a) \ar[rr]^{\hom_\CC(\CF(a),\CF(b))} & &  \CF(b) \ar[rr]^{\hom_\CC(\CF(b),\CF(c))} & & \CF(c).  
}}
\ee
Then one can see immediately that if we can physically realize the 1-functor $\CF$, it automatically realizes all linear maps $\CF_{a,b}$ for all $a,b\in\CC$ at the same time. In this sense, a 1-functor can be viewed as a parallel quantum computing.

\begin{expl}
For a given vector space $V\in\vect$, we can define two 1-functors as follows: 
\bnu
\item $-\otimes V: \vect \to \vect$ by  
$a\mapsto a \otimes V$ for an object $a\in\vect$ and by $(a\xrightarrow{f} b) \mapsto (a\otimes V \xrightarrow{f\otimes 1_V} b\otimes V)$ for a morphism $f$. 

\item $-\oplus V: \vect \to \vect$ by $a\mapsto a\oplus V$ for an object $a\in\vect$ and by $(a\xrightarrow{f} b) \mapsto (a\oplus V \xrightarrow{f\oplus 1_V} b\oplus V)$ for a morphism $f$. 
\enu
\end{expl}

\begin{rem}
A natural transformation $\phi: \CF \to \CG$ between two 1-functors $\CF, \CG: \CC \to \CD$ is a family of morphisms $\{ \phi_a: \CF(a) \to \CG(a)\}_{a\in \CC}$ such that $\phi_b\circ \CF(f) = \CG(f) \circ \phi_a$ for all  morphism $f\in \hom_\CC(a,b)$. All 1-categories (as objects), 1-functors (as 1-morphisms) and natural transformations (as 2-morphisms) form a 2-category denoted by $\mathsf{CAT}$. 
\end{rem}

\begin{rem} \label{rem:n-functor}
An $n$-functor between two $n$-categories $\CF: \CC\to \CD$ can be illustrated by the same diagram (\ref{diag:1-functor}) except that $\hom_\CC(a,b)$ is now an $(n-1)$-category and $\CF_{a,b}$ is now an $(n-1)$-functor. Then one can see that an $n$-functor realizes all $(n-1)$-functors $\CF_{a,b}$ for all $a,b\in\Ob(\CC)$ at the same time.  In this sense, an $n$-functor can be viewed as a parallel $(n-1)$-categorical computing. 
\end{rem}

A natural categorification of the one-dimensional vector space $\Cb$ is $\vect$ and a natural categorification of the $n$-dimensional vector space $\Cb^n$ is the product (or equivalently, the direct sum) $\vect^n$ of $n$ copies of $\vect$. A 1-category in the form $\vect^n$ is referred to as a finite semisimple 1-category or separable 1-category. Then the categorification of tensor product $\otimes$ of vector spaces is Deligne's tensor product $\boxtimes$ of 1-categories, i.e. $\vect^m\boxtimes\vect^n = \vect^{m n}$.
A {\em 1-categorical computation} is therefore defined to be
\[ \label{eq:1-functor}
\mbox{a 1-functor}\,\,\,\,\,F: (\vect^2)^{\boxtimes n} \to (\vect^2)^{\boxtimes n}, 
\]
where $\vect^2$ is better referred to as a {\em 1-categorical bit} (or a 1-{\it cabit}).

\begin{rem} \label{rem:n-cat-bit}
A direct generalization of 1-categorical computation is to replace the 1-categorical bit $\vect^2$ by an $n$-categorical bit (or an $n$-cabit) $n\vect^2$, where $n\vect$ is the $n$-category of $n$-vector spaces and can be obtained from $\vect$ by repeated delooping (i.e. $n\vect=\Sigma^{n-1}\vect=\Sigma^n\Cb$) \cite{JF20}. In general, there are more choices for $n$-categorical bits. For example, an $n$-categorical bit can be chosen to be a separable $n$-category, which is different from $n\vect$ in general \cite{KZ20}. 
\end{rem}


\section{Physical realization of a 1-functor} \label{sec:1-functor}
In this section, we explain that a theoretical idea of realizing a 1-functor physically by the physics of topological orders (see \cite{Wen19} for a review of topological orders and references therein). The first topological orders that were discovered in physics labs are in 2d (spatial dimension) fractional quantum Hall systems. 
Since this work only discusses a theoretical idea, we assume that physical materials that realize topological orders are abundant. Throughout this work, $n$d represents the spatial dimension.

\medskip
Consider the physical configuration of topological orders depicted in the first picture in Figure\,\ref{fig:1-functor}. It depicts a (potentially unstable) anomaly-free 1d topological order, together with a 0d boundary. By the mathematical theory of topological order (see for example \cite{KWZ15}), the 0d boundary can be mathematically described by a pair $(\CX,x)$, where $\CX$ is a separable 1-category (i.e. a finite semisimple 1-category) \footnote{More precisely, $\CX$ should be a unitary finite semisimple 1-category. We hide the unitarity requirement for convenience because the unitarity is not essential in the understanding of our ideas.} and $x$ is an object in $\CX$. Physically, this $x$ is a particle-like topological defect located at the boundary, and 1-morphisms in $\CX$ can be viewed as instantons.

According to the boundary-bulk relation \cite{KWZ15,KWZ17}, the categorical description of the bulk is the center of that of a boundary. As a consequence, 
the 1d topological order in the first picture in Figure\,\ref{fig:1-functor} can be described by the 1-category of 1-functors from $\CX$ to $\CX$, denoted by $\Fun(\CX,\CX)$. Objects $\CF,\CG$ in $\Fun(\CX,\CX)$ are particle-like topological defects in this 1d topological order. The fusion of two such defects corresponds to the composition of two 1-functors. Given such a particle-like defect, say $\CF$, if we push it to the boundary, then it fuses with the boundary particle $x$ and change it to $\CF(x)$ (see Figure\,\ref{fig:1-functor}). In other words, a 1-functor $\CF$ can be realized by a creating a particle-like defect in the 1d topological order followed by a fusion process. Creating a defect in a topological order can be physically achieved by inserting an impurity. The whole process can be simplified to inserting an impurity in the neighborhood of the boundary.

\begin{rem}
In this way, we realize any 1-functors between two different separable 1-categories $\CX_1$ and $\CX_2$ because they are just the special cases when $\CX = \CX_1 \oplus \CX_2$. 
\end{rem}

\begin{figure}
\[
\begin{array}{c}
\begin{tikzpicture}[scale=0.8]
\draw[->-,very thick] (0,3)--(0,0) node[midway,right] {$\mathrm{Fun}(\CX,\CX)$} ;
\fill (0,0) circle (0.1) node[below] {$(\CX,x)$} ;
\end{tikzpicture}
\end{array}
\quad\quad\quad 
\begin{array}{c}
\begin{tikzpicture}[scale=0.8]
\draw[->-,very thick] (0,3)--(0,1.5) node[midway,right] {$\mathrm{Fun}(\CX,\CX)$} ;
\draw[->-,very thick] (0,1.5)--(0,0) ; 
\fill (0,1.5) circle (0.1) node[left] {$\CF$} ;
\fill (0,0) circle (0.1) node[below] {$(\CX,x)$} ;
\end{tikzpicture}
\end{array}
\xrightarrow{\mbox{fusing $\CF$ with $x$}}
\begin{array}{c}
\begin{tikzpicture}[scale=0.8]
\draw[->-,very thick] (0,3)--(0,0) node[midway,right] {$\mathrm{Fun}(\CX,\CX)$} ;
\fill (0,0) circle (0.1) node[below] {$(\CX,\CF(x))$} ;
\end{tikzpicture}
\end{array}
\]
\caption{The physical realization of a 1-functor $\CF: \CX\to\CX$.}
\label{fig:1-functor}
\end{figure}

All (anomaly-free) 1d topological orders can be mathematically described by $\Fun(\CX,\CX)$ for some separable 1-category $\CX$. When $\CX\neq \vect$, $\Fun(\CX,\CX)$ is a multi-fusion 1-category instead of a fusion 1-category (see a review \cite{EGNO15}). Mathematically, it means that the identity 1-functor $\id_\CX$ is not a simple object in $\Fun(\CX,\CX)$, or equivalently, $\hom_{\Fun(\CX,\CX)}(\id_\CX,\id_\CX)\nsimeq \Cb$. Physically, it means that the multi-fusion 1-category $\Fun(\CX,\CX)$ describes an unstable phase, which can flow to a stable one if we introduce certain perturbation to the phase \cite{KWZ15}. In this case, this unstableness demands fine tuning, which makes the physical realization not fault tolerant.

\begin{figure}[htbp]
\[
\begin{array}{c}
\begin{tikzpicture}[scale=0.8]
\draw[->-,very thick] (0,3)--(0,0) node[midway,right] {$\mathrm{Fun}(\CX,\CX)$} ;
\fill (0,0) circle (0.1) node[below] {$(\CX,x)$} ;
\end{tikzpicture}
\end{array}
\xleftarrow{dimensional\,\ reduction}
\begin{array}{c}
\begin{tikzpicture}[scale=0.8]
\fill[gray!20] (0,0)--(1.5,3)--(-1.5,3)--cycle ;
\draw[->-,very thick] (0,0)--(1.5,3) node[midway,right] {$\CQ$} ;
\draw[->-,very thick] (-1.5,3)--(0,0) node[midway,left] {$\CP$} ;
\fill (0,0) circle (0.1) node[below] {$(\CX,x)$} ;
\node at (0,2) {$\CC$} ;
\end{tikzpicture}
\end{array}
\quad\quad
\begin{array}{c}
\begin{tikzpicture}[scale=0.8]
\fill[gray!20] (0,0)--(1.5,3)--(-1.5,3)--cycle ;
\draw[->-,very thick] (0,0)--(0.7,1.4) node[midway,right] {$\CQ$} ;
\draw[->-,very thick] (0.7,1.4)--(1.5,3) node[midway,right] {$\CQ$} ;
\draw[->-,very thick] (-1.5,3)--(-0.7,1.4) node[midway,left] {$\CP$} ;
\draw[->-,very thick] (-0.7,1.4)--(0,0) node[midway,left] {$\CP$} ;
\draw[->-,very thick] (-0.7,1.4)--(0.7,1.4) node[midway,above] {$\CR$} ;
\fill (0,0) circle (0.1) node[below] {$(\CX,x)$} ;
\fill (0.7,1.4) circle (0.1) node[right] {$(\CN,n)$} ;
\fill (-0.7,1.4) circle (0.1) node[left] {$(\CM,m)$} ;
\node at (0,2.5) {$\CC$} ;
\node at (0,0.8) {$\CC$} ;
\end{tikzpicture}
\end{array}
\]
\caption{These figures illustrate the idea of fixing the unstable problem in Figure\,\ref{fig:1-functor}.}
\label{fig:functor-2}
\end{figure}
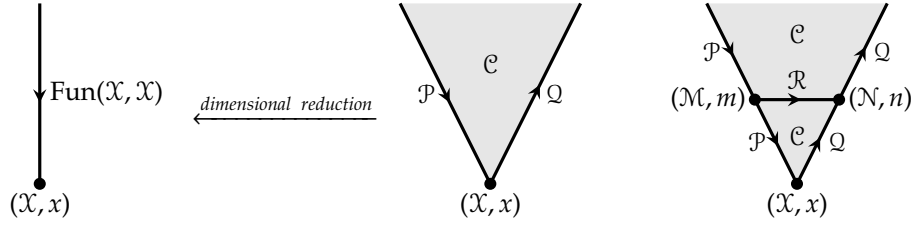

This problem can be solved by replacing the potentially unstable physical configuration depicted in the first picture in Figure\,\ref{fig:functor-2} by a stable higher dimensional physical configuration as depicted in the second picture in Figure\,\ref{fig:functor-2}.  
In this new configuration, $\CC$ labels an anomaly-free 2d topological order, and $\CP,\CQ$ label two 1d gapped boundaries of $\CC$, and the same pair $(\CX,x)$ is now realized as a domain wall between $\CP$ and $\CQ$. Mathematically, $\CC$ is a modular tensor 1-category \cite{T94}, and $\CP,\CQ$ are fusion 1-categories. All separable 1-category $\CX$ can be realized as a corner of this configuration by properly selecting $\CP,\CC,\CQ$. The second picture in Figure\,\ref{fig:functor-2} can reproduce the first picture by a dimensional reduction process (i.e. by closing the fan). After the dimensional reduction, $\Fun(\CX,\CX)$ is unstable as a 1d phase, but before the dimensional reduction, everything is stable. Then a 1-functor $\CF: \CX \to \CX$ can be realized by introducing a 0d domain wall $(\CM,m)$ in $\CP$ and a 0d wall $(\CN,n)$ in $\CQ$, and a 1d domain wall $\CR$ in $\CC$, then pushing $(\CM,m), \CR, (\CN,n)$ down to the corner (i.e. squeezing the triangle to a point) such that they fuse with $(\CX,x)$ to give $(\CX,\CF(x))$ \cite{AKZ17}. In reality, this process can be achieved by simply inserting new materials in a highly controlled way in the neighborhood of the corner directly. 
All 1-functors $\CF: \CX \to \CX$ can be physically realized in this way by properly selecting $\CM, \CR, \CN$. We believe that this way of doing computation is fault tolerant because topological defects are stable under the perturbation of local operators. 

\medskip
However, as one can see, such a `1-functor' does not realize any parallel quantum computing because only two objects $x, \CF(x) \in \CX$ appear in above process. It is not surprising because a 0d boundary is essentially a quantum mechanics system. In order to achieve the second quantization of quantum computing, we need to consider higher dimensional topological orders. A more realistic realization of 1-functor is given in the discussion of Figure\,\ref{fig:monoidal-functor} and Figure\,\ref{fig:FQH} (see also Remark\,\ref{rem:realize-1-functor}). 

\begin{rem}
A physical realization of 1-functor does not demands that of 1-category as a prerequisite because `a physical realization of 1-functors' should be viewed as a realization of 1-morphisms in a 2-category that is only equivalent to the 2-category $\mathsf{CAT}$. However, the ingredients of a 1-category $\CA$, i.e. $\{ \hom_\CA(a,b) \}_{a,b\in\CA}$ and composition maps, do have physical meanings as the spaces of instantons and the fusion of instantons, respectively (see an expository review \cite{KZha22}). The space of instantons can also be viewed as the space of ground state degeneracy (GSD) by the state-field correspondence. Unfortunately, it is not clear to us how to experimentally extract or control the information of instantons in spacetime directly. It is also not clear if it is necessary because categorical computation demands us to think about everything, including information storage and processing, not in the usual set-theoretical way by manipulating instantons but in a categorical way by manipulating functors. We leave this issue to the future. 
\end{rem}

\section{Higher categorical computation}

Above ideas generalize to higher categories and higher functors via higher dimensional topological orders but with important new features. We first generalize the discussion in Section\,\ref{sec:1-functor} to $n$-functors between $n$-categories, then we give more details on the $n=2$ case. 

\medskip
Roughly speaking, an $n$-category $\CC$ consists of a set of objects $a,b,c, \cdots$, the hom space $\hom_\CC(a,b)$ is an $(n-1)$-category, the identity 1-morphisms and the compositions of morphisms (as $(n-1)$-functors) satisfying more complicated axioms (recall Remark\,\ref{rem:n-cat}). We prefer not to give details. An $n$-functor is defined similarly to a 1-functor (recall Remark\,\ref{rem:n-functor}). In particular, an $n$-functor $\CF:\CC \to \CD$ consists of all $(n-1)$-functors $\CF_{a,b}: \hom_\CC(a,b) \to \hom_\CD(\CF(a),\CF(b))$ at the same time. Therefore, an $n$-categorical computation can be viewed as a parallel $(n-1)$-categorical computation. The first major challenge for above ideas to work is to figure out how to realize higher functors physically.  

Using the same Figure\,\ref{fig:1-functor}, now let $\CX$ be a separable $n$-category \cite[Definition\,3.3]{KZ20} and $x$ an object in $\CX$. The pair $(\CX,x)$ now describes a potentially anomalous $(n-1)$d topological order \cite{KWZ15,JF20,KZ20}. By the boundary-bulk relation, its $n$d bulk is again given by $\Fun(\CX,\CX)$, which is now a multi-fusion $n$-category. An $n$-fucntor $\CF: \CX \to \CX$ labels a topological defect of codimension 1 in the $n$d bulk, and can be realized physically in the same way as illustrated in Figure\,\ref{fig:1-functor} and Figure\,\ref{fig:functor-2}, where $\CC$ should be viewed as an $(n+1)$d bulk phase. We provide more details later for $n=2$ cases. 


\medskip
Unlike the $n=1$ case where a separable 1-categories is simply a finite product of $\vect$, a separable $n$-category does not have such a decomposition when $n\ge2$. In general, a separable $n$-category is a finite direct sum of indecomposable separable $n$-category \cite{DR18,GJF19,JF20,KZ20}. An indecomposable separable $n$-category is connected by 1-morphisms. In other words, there is only one connecting component. In general, an indecomposable separable $n$-category can have infinitely many simple objects. Its full subcategory consisting of only two objects $x$ and $y$ can be illustrated by the following diagram:
$$ 
\xymatrix{
x \ar@(ul,ur)[]^{\hom(x,x)}  \ar@/^/[rr]^{\hom(x,y)} & & y \ar@(ul,ur)[]^{\hom(y,y)} \ar@/^/[ll]^{\hom(y,x)}
}
$$
where $\hom(x,x)$ and $\hom(y,y)$ are multi-fusion $(n-1)$-categories, and $\hom(x,y), \hom(y,x)$ are separable $(n-1)$-categories. Therefore, the notion of a separable $n$-category can be defined inductively. 

\begin{expl}
We illustrate an example of an indecomposable separable 2-category $2\mathrm{Rep}(\Zb_2)$ by the following diagram \cite{DR18}. 
$$ 
\xymatrix{
x \ar@(ul,ur)[]^{\hom(x,x)}  \ar@/^/[rr]^{\hom(x,y)} & & y \ar@(ul,ur)[]^{\hom(y,y)} \ar@/^/[ll]^{\hom(y,x)}
& = & 
\one \ar@(ul,ur)[]^{\mathrm{Rep}(\Zb_2)}  \ar@/^/[rr]^{1\vect} & & \one_c \ar@(ul,ur)[]^{1\vect_{\Zb_2}} \ar@/^/[ll]^{1\vect}
}
$$
where $x=\one$ and $y=\one_c$ denote the only two simple objects in $2\mathrm{Rep}(\Zb_2)$, and $\hom(\one,\one)=\mathrm{Rep}(\Zb_2)$ is the 1-category of finite dimensional representations of the $\Zb_2$-group, and $\hom(\one_c,\one_c)=\vect_{\Zb_2}$ is the 1-category of $\Zb_2$-graded vector spaces. All other objects are finite direct sum of $\one$ and $\one_c$. The separable 2-category $2\mathrm{Rep}(\Zb_2)$ has a physical meaning in 3d toric code model or finite gauge theory with the gauge group $G=\Zb_2$ \cite{KTZ20}. 
\end{expl}

\begin{rem}
The classification of separable $n$-categories is not known. It was known that a separable $n$-category is the representation category of a multi-fusion $(n-1)$-category \cite{DR18,JF20,KZ20}. Unfortunately, even the classification of fusion 1-categories is not known.
\end{rem}

Therefore, when $n\ge2$, there is no canonical choice of an $n$-categorical bit. The simpest choice is $n\vect^2$ (recall Remark\,\ref{rem:n-cat-bit}), but this choice does not exhibit the richness of $n$-categorical computation.  An {\em $n$-categorical computation} is then should be defined as
\[ \label{eq:n-functor}
\mbox{an $n$-functor}\,\,\,\,\,F: \CX \to \CX, 
\]
where $\CX$ is a separable $n$-category. 

Importantly, the data $(\CX,x)$ is essentially equivalent to the multi-fusion $(n-1)$-category $\hom_\CX(x,x)$ \cite{JF20,KZ20}. The above physical realization of an $n$-functor $\CF: \CX \to \CX$ can also be viewed as a physical realization of a monoidal $(n-1)$-functor from $\hom_\CX(x,x)$ to $\hom_\CX(\CF(x),\CF(x))$. This physical realization of a monoidal $(n-1)$-functor can be stated as a precise mathematical theorem (see  \cite[Theorem\ 3.2.3]{KZ18} for $n=2$ and \cite{KZ21} for general $n$).

\begin{rem}
Topological orders were proposed long ago to provide the physical realization of the fault tolerant quantum computation \cite{K03,F98} (see for example \cite{Wan10} for a review and references therein). Our proposal suggests that the physics of topological orders might allows us to do $n$-categorical computations in a fault tolerant way. 
\end{rem}

Although the physical realization of topological orders cannot go beyond 3d, it is already very interesting and rich in 1d and 2d cases because anomalous 1d topological orders and 2d topological orders are abundant. In the rest of this paper, we further illustrate the idea for anomalous 1d topological orders because it is experimentally realizable either as a gapped domain wall in a 2d fractional quantum Hall system or a gapped boundary in a toric code model achieved in quantum simulation (see for example \cite{S22,SLK+21} and references therein). 

\begin{figure}[htbp]
\[
\begin{tikzpicture}[scale=0.8]
\fill[gray!30] (-3,0) rectangle (3,3) ;
\draw[very thick] (3,0)--(3,3) ; 
\draw[very thick] (1,0)--(1,3) ;
\draw[fill=white] (2.9,0.4) rectangle (3.1,0.6) node[midway,right] {$a_n$} ;
\draw[fill=white] (2.9,2.4) rectangle (3.1,2.6) node[midway,right] {$a_1$} ;
\draw[fill=white] (2.9,1.9) rectangle (3.1,2.1) node[midway,right] {$a_2$} ;
\node at (3,3.3) {$\CA$} ;
\node at (1,3.3) {$\CF$} ;
\node at (2,1.5) {$\FZ_1(\CA)$} ;
\node at (-1,1.5) {$\CC$} ;
\node at (3.2,1.4) {$\cdot$} ;
\node at (3.2,1.2) {$\cdot$} ;
\node at (3.2,1.0) {$\cdot$} ;
\draw[decorate,decoration=brace,very thick] (3,-0.2)--(1,-0.2) ;
\node at (3,-0.7) {$\CF\boxtimes_{\FZ_1(\CA)} \CA=\CB$} ;
\end{tikzpicture}
\]
\caption{the idea of physically realizing a monoidal 1-functor}
\label{fig:monoidal-functor}
\end{figure}
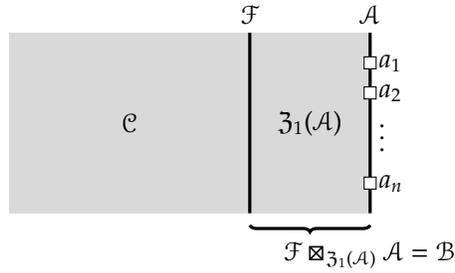

An anomalous 1d topological order can be described by a pair $(\CX,x)$, where $\CX$ is a separable 2-category and $x\in \CX$ is a distinguished object\footnote{Without loss of generality, we can assume $x$ is indecomposable.}. Equivalently, it can also be described a fusion 1-category $\CA:=\hom_\CX(x,x)$, the objects of which are particle-like topological excitations (or topological defects). We illustrate the idea of physically realizing a monoidal 1-functor $\CF: \CA \to \CB$ from $\CA$ to another multi-fusion 1-category $\CB$ in Figure\,\ref{fig:monoidal-functor}. 

Since the 1d topological order $\CA$ is anomalous, it can only be a gapped boundary of a 2d topological order, whose particle-like topological excitations form a braided fusion 1-category given by the Drinfeld center of $\CA$ \cite{KK12,KWZ15,KWZ17}, denoted by $\FZ_1(\CA)$. Consider a gapped domain wall $\CF$ between the bulk of $\CA$ and the anomaly-free 2d phase $\CC$. The particle-like topological excitations on the wall form a multi-fusion 1-category still denoted by $\CF$.

Topological excitations $a_1, \cdots, a_n$ in the anomalous 1d topological order can be arranged to be separated with equal distance. According to \cite[Theorem\ 3.2.3]{KZ18}, the gapped domain wall or the fusion category $\CF$ determines (actually is equivalent to) a monoidal functor from $\CA$ to $\CB:=\CF\boxtimes_{\FZ_1(\CA)} \CA$ defined by 
$$
a \mapsto \one_\CF \boxtimes_{\FZ_1(\CA)} a \in \CF\boxtimes_{\FZ_1(\CA)} \CA=\CB, \quad\quad \forall a\in \CA,
$$
where $\one_\CF$ denotes the trivial particle (or the tensor unit) in $\CF$. In other words, a physical realization of above functor can be achieved by (1) fusing a 2d phase with a gapped boundary onto 
the bulk of $\CA$ such that a new 2d phase $\CC$ and a gapped domain wall $\CF$ between $\CC$ and the bulk of $\CA$ are created; (2) fusing $\CF$ with $\CA$.\footnote{One way to realize ``fusing $\CF$ with $\CA$'' is to create $\CF$ in a neighborhood of $\CA$.} In a special case $\CC$ coincides with the bulk of $\CA$, one can replace the step (1) by selecting a line in the bulk of $\CA$ then modifying the microscopic physics along the line to create a gapped domain wall $\CF$ directly. Note that this modification might be achieved by a macroscopic process (e.g. gluing new materials or chemicals along the line). Since both $\CC$ and $\CF$ are many-body systems, this type of computation need manipulate infinitely degrees of freedom in the thermodynamics limit and is thus a second quantized computation. 

\begin{rem} \label{rem:realize-1-functor}
If we fix the location of $a_i\in \CA$ as in Figure\,\ref{fig:monoidal-functor}, we can ignore the monoidal structure on $\CA$ and view $\CF$ as a physical realization of a 1-functor. 
\end{rem}

For example, it is possible to control and arrange anyons in 2d in the experiments of fractional quantum Hall systems (see for example \cite{NLGM22} and references therein). As depicted in Figure\,\ref{fig:FQH}, by simply arranging these anyons in fractional quantum Hall systems along the dashed line in Figure\,\ref{fig:FQH}, we obtain a realization of $\CA$ in Figure\,\ref{fig:monoidal-functor}, where $\FZ_1(\CA)=\CA\boxtimes\overline{\CA}$ is now a double layered fractional quantum Hall system. An example of $\CF$ is realized by two 1d domain walls $\CF'$ and $\CF''$ sitting on the two sides of the dashed line as depicted in Figure\,\ref{fig:FQH}, i.e. $\CF=\CF'\boxtimes \CF''$.
\begin{figure}[htbp]
\[
\begin{tikzpicture}[scale=0.8]
\fill[gray!30] (-3,0) rectangle (3,3) ;
\draw[dashed] (0,0)--(0,3) ; 
\draw[very thick] (1,0)--(1,3) ;
\draw[very thick] (-1,0)--(-1,3) ;
\draw[fill=white] (-0.1,0.4) rectangle (0.1,0.6) node[midway,right] {$a_n$} ;
\draw[fill=white] (-0.1,2.4) rectangle (0.1,2.6) node[midway,right] {$a_1$} ;
\draw[fill=white] (-0.1,1.9) rectangle (0.1,2.1) node[midway,right] {$a_2$} ;
\node at (0,3.3) {$\CA$} ;
\node at (-1,3.3) {$\CF'$} ;
\node at (1,3.3) {$\CF'$} ;
\node at (-2,1.5) {$\CA$} ;
\node at (2,1.5) {$\CA$} ;
\node at (-0.5,1.5) {$\CA$} ;
\node at (0.2,1.4) {$\cdot$} ;
\node at (0.2,1.2) {$\cdot$} ;
\node at (0.2,1.0) {$\cdot$} ;
\draw[decorate,decoration=brace,very thick] (1,-0.2)--(-1,-0.2) ;
\node at (0,-0.7) {$\CF\boxtimes_{\FZ_1(\CA)} \CA=\CF'\boxtimes_\CA \CA \boxtimes_\CA \CF'' = \CB$} ;
\end{tikzpicture}
\]
\caption{the idea of realizing a monoidal functor $\CF: \CA \to \CB$ defined by $a \mapsto \one_{\CF'} \boxtimes_\CA a \boxtimes_\CA \one_{\CF''} \in \CB$ in a 2d fractional quantum Hall system}
\label{fig:FQH}
\end{figure}

\begin{rem}
The idea illustrated in Figure\,\ref{fig:monoidal-functor} and Figure\,\ref{fig:FQH} can be automatically generalized to 3d topological orders to give a physical realization of monoidal 2-functors. 
\end{rem}

The theoretical idea we present here is still far from an experimental realization. 
At this stage, it is too early to tell if the categorical computation is technologically possible or impossible. It depends on the future development of the physics of topological phases, the discovery of new topological materials and the technology advances in controlling and engineering topological defects. We also do not know if it can be more efficient than classical/quantum computation. However, we believe that this theoretical idea deserves some attentions from theorists who are working closely with experimentalists. Moreover, theoretically, it seems rather obvious that categorical computation is suitable or very likely to be powerful in computing (higher) categories and (higher) functors. On the other hand, it is not clear how to simulate a category or a functor by a classical computer or a quantum computer. 

We believe that it is worthwhile to make this naive idea available to experts so that more ideas and discussion can follow.  For example, many natural questions can be asked based on the proposal of this work, such as the details of the physical realization, in what sense the computation can be made `universal', what possible gates are, what the complexity theory of categorical computation is, develop algorithms for problems in category theory such as how to factorize a modular tensor category into primary ones, etc.

\begin{rem}
We want to remind readers that the higher categorical computation is fundamentally different from the quantum computing based on the braidings between anyons or higher dimensional topological defects in higher dimensional topological orders. Both fusion matrices and braiding matrices for higher dimensional topological defects are the defining data of certain natural transformations between higher functors. Therefore, it might be possible to use fusions or braidings of higher dimensional topological defects to simulate or compute certain categorical structures. It is very interesting to explore the relation or possible interaction between these two different ideas of computing. We will leave it to the future. 
\end{rem}

\bigskip
\noindent{\bf Acknowledgement}: We thank Zheng-Wei Liu, Ce Shen, Xiao-Ming Sun, Zhong Wang and Bo Yang for comments. We are supported by Guangdong Provincial Key Laboratory (Grant No.2019B121203002). LK is also supported by NSFC under Grant No. 11971219 and by Guangdong Basic and Applied Basic Research Foundation under Grant No. 2020B1515120100. HZ is also supported by NSFC under Grant No. 11871078.

\end{document}